\documentclass[prd,nofootinbib,preprint]{revtex4}

\usepackage{amsmath,amssymb}
\usepackage{epsfig}

\newcommand{\sla}[1]{/\!\!\!#1}

\begin{document}

\preprint{\hbox{YITP-SB-06-10}} 
\preprint{\hbox{IFUSP-1620/2006}}

\title{$pp\rightarrow j j e^\pm \mu^\pm \nu \nu$ and $j j e^\pm \mu^\mp \nu
\nu$ at ${\cal O}(\alpha_{\rm em}^6)$ and ${\cal O}(\alpha_{\rm em}^4
\alpha_s^2)$ for the Study of the Quartic Electroweak Gauge Boson Vertex at
LHC }

\author{O.\ J.\ P.\ \'Eboli$^1$}
\email{eboli@fma.if.usp.br}

\author{M.\ C.\ Gonzalez-Garcia$^2$}
\email{concha@insti.physics.sunysb.edu}

\author{J.\ K.\  Mizukoshi$^1$}
\email{mizuka@fma.if.usp.br}

\affiliation{$^1$ Instituto de F\'{\i}sica, 
Universidade de S\~ao Paulo, S\~ao Paulo -- SP, Brazil.}

\affiliation{$^2$ Y.I.T.P., SUNY at Stony Brook, Stony Brook, NY
  11794-3840, USA}

\affiliation{
  IFIC, Universitat de Val\`encia - C.S.I.C., Apt 22085, 46071
  Val\`encia, Spain.\\}

\vspace*{1.5cm}
\begin{abstract}
  \vspace*{.5cm} 
  We analyze the potential of the CERN Large Hadron
  Collider (LHC) to study the structure of quartic vector--boson
  interactions through the pair production of electroweak gauge bosons
  via weak boson fusion $ q q\to q q WW$.  In order to study these
  couplings we have performed a partonic level calculation of all
  processes $pp \rightarrow j j e^\pm \mu^\pm \nu \nu$ and
  $pp\rightarrow j j e^\pm \mu^\mp \nu \nu$ at the LHC using the exact
  matrix elements at ${\cal O}(\alpha_{\rm em}^6)$ and ${\cal
    O}(\alpha_{\rm em}^4 \alpha_s^2)$ as well as a full simulation of
  the $t \bar{t}$ plus 0 to 2 jets backgrounds.  A complete
  calculation of the scattering amplitudes is necessary not only for a
  correct description of the process but also to preserve all
  correlations between the final state particles which can be used to
  enhance the signal.  Our analyses indicate that the LHC can improve
  by more than one order of magnitude the bounds arising at present
  from indirect measurements.
\end{abstract}

\maketitle

\section{Introduction}

Within the framework of the Standard Model (SM), the structure of the
trilinear and quartic vector boson couplings is completely determined by the
$SU(2)_L \times U(1)_Y$ gauge symmetry. The study of these interactions can
either lead to an additional confirmation of the model or give some hint on
the existence of new phenomena at a higher scale \cite{anomalous}.  The triple
gauge--boson couplings were probed at the LEP \cite{lep,exp:LEP} and are still
under scrutiny at the Tevatron \cite{teva} through the production of vector
boson pairs. However, we have only started to study directly the quartic
gauge--boson couplings \cite{exp:LEP}. If any deviation from the SM
predictions is observed, independent tests of the triple and quartic
gauge--boson couplings can give important information on the type of new
physics responsible for the departures from the SM. For example, the exchange
of heavy bosons can generate a tree level contribution to four gauge--boson
couplings while its effect in the triple--gauge vertex would only appear at
one--loop level, and consequently be suppressed with respect to the quartic one
\cite{aew}.

At present the scarce experimental information on quartic anomalous couplings
arises from the processes $e^+e^-\rightarrow W^+W^-\gamma$, $ Z\gamma\gamma$,
$ Z Z \gamma$, and $ \nu \bar{\nu} \gamma \gamma$ at LEP~\cite{exp:LEP}.  Due
to phase space limitations, the best sensitivity is attainable for couplings
involving photons which should appear in the final state.  Photonic quartic
anomalous couplings can also affect $\gamma\gamma Z$ and $\gamma\gamma W$
productions at Tevatron~\cite{our:quartic,stir2} and they will be further
tested at LHC~\cite{our:quartic2} and in the long term at the 
next generation $e^+e^-$ collider~\cite{Belanger:1999aw,nlc,bel:bou,ggnos,our:vvv}.

Purely electroweak quartic couplings $W^+ W^- W^+ W^-$ and $W^+ W^- Z Z$ have
not been directly tested so far but will be within reach at
LHC~\cite{bagger0,bagger,dobado,Belyaev:1998ih}.  In this work we study the
potential of the LHC to probe them by performing a detailed analysis of the
most sensitive channels that are the production via weak boson fusion (WBF) of
$W^+ W^-$ pairs accompanied by jets, {\em i.e.},
\begin{equation}
p + p \to j j W^+ W^- \to j  j  e^\pm \mu^\mp \nu \nu \; ,
\label{jj}
\end{equation}
and the WBF production of a pair of jets plus $W^\pm W^\pm$
\begin{equation}
p + p \to j  j W^\pm W^\pm \to j  j e^\pm \mu^\pm \nu \nu  \; .
\label{ll}
\end{equation}
We have only considered final state with different flavor leptons ($e$ and
$\mu$) in order to avoid backgrounds coming from $Z,\gamma \to e^+ e^-$ or
$\mu^+ \mu^-$.  The advantage of WBF, where the scattered final--state quarks
receive significant transverse momentum and are observed in the detector as
far-forward/backward jets, is the strong reduction of QCD backgrounds due to
the kinematic configuration of the colored part of the event.

There are previous studies of the quartic gauge boson couplings at the
LHC.  The earlier works~\cite{bagger0,bagger,dobado} relied upon the
equivalence theorem \cite{equ} or/and the effective $W$--boson
approximation \cite{evb}.  In Ref.~\cite{Belyaev:1998ih} the full tree
level calculation of the processes $pp \to V V +$ 2 jets, with $V =
W^\pm, Z^0$ was presented.  Here, we improve over these earlier works
by computing the full matrix element for all processes with the six
fermion final states in (\ref{jj}) and (\ref{ll}) at ${\cal
  O}(\alpha_{\rm em}^6)$ and ${\cal O}(\alpha_{\rm em}^4 \alpha_s^2)$.
This includes the contribution from the resonant gauge boson pair
production considered in Ref.~\cite{Belyaev:1998ih} as well as all the
non--resonant contributions and their interference.  We have also
performed a full simulation of the $t \bar{t}$ background and
evaluated the $t \bar{t}$ plus 1 and 2 jets backgrounds using the
narrow width approximation for the top.

The interactions responsible for the electroweak symmetry breaking
play an important role in the gauge--boson scattering at high energies
as they are an essential ingredient to avoid unitarity violation in
the scattering amplitudes of massive vector bosons at the TeV scale
\cite{unit}. There are two possible forms of electroweak symmetry
breaking which lead to different solutions to the unitarity problem:
$(a)$ there is a particle lighter than 1 TeV, the Higgs boson in the
standard model, or $(b)$ such particle is absent and the longitudinal
components of the $W$ and $Z$ bosons become strongly interacting at
high energies. In the latter case, the symmetry breaking occurs due to
the nonzero vacuum expectation value of some composite operators which
are related with new underlying physics.

We parameterize in a model independent form the possible deviations of the SM
predictions for the $W^+ W^- W^+ W^-$ and $W^+ W^- Z Z$ quartic gauge
couplings in these two different scenarios as described in
Sec.~\ref{sec:theory}. In the first case we assume the existence of a light
Higgs boson and consequently we are lead to dimension eight effective
operators where the $SU(2)_L \times U(1)_Y$ gauge invariance is realized
linearly. We also contemplate the scenario where no new heavy resonance has
been observed that leads to the gauge symmetry being realized nonlinearly by
using the chiral Lagrangian approach.

Valuable information on the possibility of new physics effects can also be
gathered from the low energy data and the results of the $Z$ physics; see
Ref.~\cite{Barbieri:2004qk} for a recent review. In particular they can
constrain the possible deviations of the quartic gauge boson self--interactions
from the SM predictions through their contributions to the electroweak
radiative corrections \cite{Brunstein:1996fz}. For completeness we present in
Sec.~\ref{sec:lepbounds} the updated bounds on these effects from the global
electroweak fit.

Sections~\ref{sec:tools} and~\ref{sec:sigback} contain the details of the
strategies proposed to reduce the backgrounds to acceptable levels while
keeping the signal from the quartic gauge vertex. We find that the complete
calculation of the scattering amplitudes is necessary to preserve all
correlations between the final state particles which can be used to enhance
the signal. We also study the precision with which the background rate in
the search region can be predicted which is the ultimately limiting factor.

Our final quantitative results on the attainable sensitivity at LHC
are presented in Sec.~\ref{sec:results}. We find that LHC can improve
by more than one order of magnitude the bounds arising at present from 
indirect measurements and it is able to test deviations with the size 
expected in the scenario in which no light Higgs boson is found and 
the gauge symmetry is realized nonlinearly.

\section{Theoretical framework}
\label{sec:theory}

In this work we focus on the study of the structure of the weak quartic
couplings containing $W^\pm$'s and/or $Z$'s. For the sake of simplicity we
will consider effective interactions that do not contain derivatives of the
gauge fields. With this requirement there are only two possible Lorentz
invariant structures contributing to each of the four gauge boson vertices
\begin{eqnarray}
 && {\cal O}^{WW}_0= g^{\alpha \beta} g^{\gamma \delta} 
\left [ W^+_\alpha  W^-_\beta W^+_\gamma W^-_\delta\right] \; , 
\;\;\;\;\;\;\;\;\;  
  {\cal O}^{WW}_1 = g^{\alpha \beta} g^{\gamma \delta} \left [ 
   W^+_\alpha  W^+_\beta W^-_\gamma W^-_\delta \right] \; ,
\nonumber \\
&&  {\cal O}^{WZ}_0= g^{\alpha \beta} g^{\gamma \delta} 
\left [ W^+_\alpha  Z_\beta W^-_\gamma Z_\delta\right] \; , 
\;\;\;\;\;\;\;\;\;\;\;\;   \;\;   
  {\cal O}^{WZ}_1 = g^{\alpha \beta} g^{\gamma \delta} \left [ 
W^+_\alpha  W^-_\beta Z_\gamma Z_\delta \right] \; ,
\label{lor} \\
&& {\cal O}^{ZZ}_0={\cal O}^{ZZ}_1\equiv {\cal O}^{ZZ} =
g^{\alpha \beta} g^{\gamma \delta} \left [ 
Z_\alpha Z_\beta Z_\gamma Z_\delta \right ] \; ,
\nonumber 
\end{eqnarray}
and the Lagrangian for the four gauge boson vertex will be  
\begin{equation}
{\cal L}^{VVV'V'} \equiv\, c^{VV'}_0\, {\cal O}^{VV'}_0\,+\, c_1^{VV'}\, 
{\cal O}^{VV'}_1 \: . 
\label{eq:lfour}
\end{equation}

In the SM, $SU(2)_L$ gauge invariance and renormalizability imply that
\begin{equation}
c_{0,\rm SM}^{WW}=-c_{1,\rm SM}^{WW}=\frac{2}{c_W^2} c_{0,\rm SM}^{WZ}
=-\frac{2}{c_W^2}c_{1,\rm SM}^{WZ}=g^2 \;\;\;\;\;\;\;\; 
c^{ZZ}_{\rm SM} =0
\label{eq:smcoeff}
\end{equation}
where $c_W$  is the cosine of the weak mixing angle and 
$g$ is the $SU(2)_L$ coupling constant. 

Conversely, if the SM is thought of only as an effective low energy theory
valid up to the scale $\Lambda$, one expects deviations from Eq.\
(\ref{eq:smcoeff}) even if we still retain the gauge symmetry group, the
fermionic spectrum, and the pattern of spontaneous symmetry breaking (EWSB) as
valid ingredients to describe Nature at energies $E\ll\Lambda$. In this case
one can still write the Lagrangian for the four gauge boson interactions as
Eq.\ (\ref{eq:lfour}) but now the coefficients, $c_0$ and $c_1$ will be in
general independent, and we can write
\begin{equation}
c^{VV'}_i=c_{i, \rm SM}^{VV'}\, +\, g^2 \, \Delta c^{VV'}_i\; . 
\label{eq:anocoeff}
\end{equation}
In the language of effective Lagrangians the deviations $\Delta c_i$
will be generated by higher dimension operators parameterizing the low
energy effect of the new physics. The order on the expansion at which
these deviations are expected to appear depends on whether the low
energy spectrum still contains a light SM--like Higgs boson
responsible of EWSB or, on the contrary, EWSB is due to a heavy (or
not fundamental) Higgs boson.

\subsection{Effective Operators with Linear Realization 
of the $SU(2)_L\times U(1)$ Gauge Symmetry}

We first assume that the low energy spectrum contains a light Higgs
boson. In this case we chose a linear realization of the symmetry
breaking in the form of the conventional Higgs doublet field $\Phi$.
In the usual effective Lagrangian language, at low energy we describe
the effects of the new physics -- which will manifest itself directly
only at scales above $\Lambda$ -- by including higher--dimension
operators in the Lagrangian.  The basic blocks for constructing the
operators which can modify the four gauge boson electroweak vertices
are the Higgs field, its covariant derivative $D_\mu \Phi$, the
$SU(2)_L$ field strength $W^i_{\mu\nu}$, and $U(1)_Y$ field strength
$B_{\mu\nu}$.  The lowest order operators which can be built are of
dimension six~\cite{linear}. However dimension six operators which
modify the four gauge boson vertices, affect either the two or three
gauge boson couplings as well.  Consequently they are better searched
for, and severely constrained at present, by looking into those
effects.

The lowest dimension operators that modify the quartic boson
interactions but do not exhibit two or three weak gauge boson vertices
are dimension 8. The counting is straight forward: one can get a weak
boson field either from the covariant derivative of $\Phi$ or from the
field strength tensor. In either case the vector field is either
accompanied by a vacuum expectation value (VEV) of the Higgs field
($v$) or a derivative.  Therefore genuine quartic vertices are of
dimension 8 or higher. There are only two independent dimension 8
operators without derivatives of the gauge fields (for further details
see appendix~\ref{app:dim8})
\begin{eqnarray}
  {\cal L}_{S,0} &=& \frac{f_0}{\Lambda^4} 
\left [ \left ( D_\mu \Phi \right)^\dagger
 D_\nu \Phi \right ] \times
\left [ \left ( D^\mu \Phi \right)^\dagger
D^\nu \Phi \right ] \; ,
\label{eq:opf0}\\
  {\cal L}_{S,1} &=& \frac{f_1}{\Lambda^4} 
\left [ \left ( D_\mu \Phi \right)^\dagger
 D^\mu \Phi  \right ] \times
\left [ \left ( D_\nu \Phi \right)^\dagger
D^\nu \Phi \right ] \; .
\label{eq:opf1}
\end{eqnarray}
When the Higgs field $\Phi$ is replaced by its VEV, (\ref{eq:opf0}) and
(\ref{eq:opf1}) generate four gauge boson interactions as Eqs.\
(\ref{eq:lfour}) and (\ref{eq:anocoeff}) with
\begin{eqnarray}
\Delta c^{WW}_i&=&\frac{g^2 v^4 f_i}{8 \Lambda^4}\equiv 
\Delta c_{i,\rm lin} 
\;, \nonumber \\
\Delta c^{WZ}_i&=&\frac{g^2 v^4 f_i}{16 c_W^2 \Lambda^4}=
\frac{\Delta c_{i,\rm lin}}{2 c_W^2}
  \;, \label{eq:dclin}\\
\Delta c^{ZZ}&=&\frac{g^2 v^2 (f_0+f_1)}{32 c_W^4 \Lambda^4}=
\frac{\Delta c_{0,\rm lin}+\Delta c_{1,\rm lin}}{4 c_W^4} \;. \nonumber
\end{eqnarray}

\subsection{Effective Operators with Non-Linear Realization 
of the $SU(2)_L\times U(1)$ Gauge Symmetry}

If the electroweak symmetry breaking is due to a heavy (strongly interacting)
Higgs boson, which can be effectively removed from the physical low-energy
spectrum, or to no fundamental Higgs scalar at all, one is led to consider the
most general effective Lagrangian which employs a nonlinear representation of
the spontaneously broken $SU(2)_L \otimes U(1)_Y$ gauge symmetry
\cite{Appelquist}. The resulting chiral Lagrangian is a non-renormalizable
non-linear $\sigma$ model coupled in a gauge-invariant way to the Yang-Mills
theory.  This model independent approach incorporates by construction the
low-energy theorems \cite{cgg}, that predict the general behavior of Goldstone
boson amplitudes irrespective of the details of the symmetry breaking
mechanism. Notwithstanding, unitarity implies that this low-energy effective
theory should be valid up to some energy scale smaller than $4\pi v \simeq 3$
TeV, where new physics would come into play.

To specify the effective Lagrangian one must first fix the symmetry breaking
pattern. We consider that the system presents a global $SU(2)_L \otimes
SU(2)_R$ symmetry that is broken to $SU(2)_C$. With this choice, the building
block\footnote{We follow the notation of Ref.~\cite{Appelquist}.} of the
chiral Lagrangian is the dimensionless unimodular matrix field $\Sigma(x)$,
which transforms under $SU(2)_L \otimes SU(2)_R$ as $(2,2)$:
\begin{equation}
\Sigma(x) ~=~ \exp\left(i \frac{\varphi^a(x) \tau^a}{v}\right) \; ,
\end{equation}
where the $\varphi^a$ fields are the would-be Goldstone fields and
$\tau^a$ ($a=1$, $2$, $3$) are the Pauli matrices.  The $SU(2)_L
\otimes U(1)_Y$ covariant derivative of $\Sigma$ is defined as
\begin{equation}
D_\mu \Sigma ~\equiv~ \partial_\mu \Sigma
+ i g \frac{\tau^a}{2} W^a_\mu \Sigma -
i g^\prime \Sigma \frac{\tau^3}{2} B_\mu \; .
\end{equation}

Quartic vector boson interactions are generated at second order ($p^4$) in the
derivative expansion \cite{Appelquist}. For simplicity we will consider only
interactions which respect the custodial $SU(2)$ symmetry.  At this order,
there are only two such operators usually denoted as
\begin{eqnarray}
{\cal L}^{(4)}_4 &=& \alpha_4\left[{\rm{Tr}}
\left(V_{\mu}V_{\nu}\right)\right]^2
\label{eq:opal4}
\;, \\
{\cal L}^{(4)}_5 &=& \alpha_5\left[{\rm{Tr}}
\left(V_{\mu}V^{\mu}\right)\right]^2
\;,
\label{eq:opal5}
\end{eqnarray}
where we defined $V_\mu \equiv \left ( D_\mu \Sigma \right ) \Sigma^\dagger$.
These effective operators generate four gauge boson interactions as Eqs.\
(\ref{eq:lfour}) and (\ref{eq:anocoeff}) with
\begin{eqnarray}
\Delta c^{WW}_0&=&g^2 \alpha_4\equiv \Delta c_{0,\rm no-lin}  
\;\;\;\;\;\;\;\;\;  \;\;\;
\Delta c^{WW}_1=g^2 \alpha_5\equiv \Delta c_{1,\rm no-lin}     
\nonumber \\
\Delta c^{WZ}_0&=&\frac{g^2 }{2 c_W^2} \alpha_4 =
\frac{\Delta c_{0,\rm no-lin}}{2 c_W^2}     
 \;\;\;\;\;\;\;\;\; 
\Delta c^{WZ}_1=\frac{g^2 }{ 2 c_W^2} \alpha_5=
\frac{\Delta c_{1,\rm no-lin}}{2 c_W^2}      \label{eq:dcnolin}   \\ 
\Delta c^{ZZ}&=&\frac{g^2}{4 c_W^4} (\alpha_4+\alpha_5)=
\frac{\Delta c_{0,\rm no-lin}+\Delta c_{1,\rm no-lin}}{4 c_W^4}     
 \;\;\;\;\;\;\;\;\; 
\nonumber
\end{eqnarray}

\section{Low energy constraints}
\label{sec:lepbounds}

Valuable information on the possibility of new physics effects can also be
gathered from electroweak precision data, measured mainly at the $Z$-peak by
LEP1 experiments, but also including the $W$ and top masses and other
measurements.  These data can be used to constrain the possible deviations of
the quartic gauge boson self-interactions from the SM predictions as they
contribute to the gauge boson self-energies at the one-loop level
\cite{Brunstein:1996fz}.

Standard Model electroweak radiative corrections as well as 
universal new physics effects enter in the predictions of these 
electroweak precision observables in three different combinations 
usually named $\varepsilon_1,\varepsilon_2,\varepsilon_3$~\cite{abc}
(or $S$, $T$, and $U$~\cite{pt}), so in general
\begin{equation}
\varepsilon_i=\varepsilon_{i,{\rm SM}}+\varepsilon_{i,{\rm new}}\; .
\end{equation}

Technically the procedure to obtain the contribution from the
operators (\ref{eq:opf0}), (\ref{eq:opf1}), (\ref{eq:opal4}), and
(\ref{eq:opal5}) to the $\varepsilon$'s is the following: first we
evaluate their contribution to the self--energies using dimensional
regularization. Then, we keep only the leading non--analytic
contributions -- that is, the terms proportional to log$(\mu^2)$ --
dropping all others. These contributions are easily obtained by the
substitution
\begin{equation}
\frac{2}{4-d} \rightarrow {\rm{log}}\;\frac{\Lambda^2}{M_Z^2}\; ,
\nonumber
\end{equation}
where $\Lambda$ is the energy scale which characterizes the appearance
of new physics.

With this procedure we found in Ref.~\cite{Brunstein:1996fz} that for
the operators (\ref{eq:opf0}), (\ref{eq:opf1}), (\ref{eq:opal4}), and
(\ref{eq:opal5}), $\varepsilon_{2, \rm new} = \varepsilon_{3, \rm new}
= 0$ and that only $\varepsilon_{1, \rm new}$ is non-vanishing:
\begin{eqnarray}
\varepsilon_{1, \rm new}= & - 
\frac{\displaystyle 15 g^2 \Delta c_0}{\displaystyle 64\pi^2} 
(1+c_W^2)\frac{s_W^2}{c_W^2}
{\rm{log}}\;\frac{\Lambda^2}{M_Z^2}
\; , 
\label{e1:0}
\\
\varepsilon_{1, \rm new}= & - \frac{\displaystyle 3 g^2 \Delta c_1
}{\displaystyle 32 \pi^2} 
(1+c_W^2)\frac{s_W^2}{c_W^2}
\; {\rm{log}}\;\frac{\Lambda^2}{M_Z^2}
\; ,
\label{e1:1}
\end{eqnarray}
where $\Delta c_i$ for the case of linear [non-linear] realization of
the gauge symmetry are defined in Eq.~(\ref{eq:dclin})
[Eq.~(\ref{eq:dcnolin})]

Recent global analysis of the low energy and LEP data 
\cite{Barbieri:2004qk} yields 
\[
   \varepsilon_1  = (5.0 \pm 1.1) \times 10^{-3} \; ,
\]
while the SM prediction is a function of $m_t$, $m_h$, $\alpha_s$ and 
$\alpha_{\rm em}$. We use $m_t = 178 $ GeV, 
$\alpha_s(M_Z) = 0.119$ and  $\alpha_{\rm em}(M_Z) = 1/128.88$.  

For the case with a  light Higgs boson of $m_h=120$ GeV 
and  a new physics scale $\Lambda= 2$ TeV we find that at 99\% CL 
\begin{eqnarray}
   -5.2 < &f_0\times 10^{-3}& <9.0 \; ,
\nonumber \\
 -13 < &f_1\times 10^{-3} & < 22 \; .
\label{eq:lel}
\end{eqnarray} 

In models without a light Higgs boson, the gauge-boson contribution to
$\varepsilon_1$ is infinite as a consequence of the absence of the
elementary Higgs. On the other hand, we must also include the tree
level effect due to the ${ \cal O}(p^4)$ operator which violates
custodial $SU(2)$ and which absorbs this infinity through the
renormalization of the corresponding coefficient.  If the
renormalization condition is imposed at a scale $\Lambda$, we are left
with the contribution due to the running from the scale $\Lambda$ to
$M_Z$. Therefore, the SM contribution without the Higgs boson will be
the same as that of the SM with an elementary Higgs boson, with the
substitution $\ln (M_H)\rightarrow \ln(\Lambda)$. For $\Lambda= 2$ TeV
we get the following 99\% CL bounds
\begin{eqnarray}
 -0.32 < &\alpha_4& <0.085 \; ,
%
\nonumber \\
-0.81 < &\alpha_5& <0.21 \; .
\label{eq:lenl}
\end{eqnarray}

\section{Calculation tools}
\label{sec:tools}

We concentrate on the study of the structure of quartic vector--boson
interactions through the production of $W^\pm W^\mp$ and $W^\pm W^\pm$
in WBF, with subsequent decay to $e\mu$ pairs and neutrinos.  The
signal is thus characterized by two quark jets, which typically enter
in the forward and backward regions of the detector and are widely
separated in pseudorapidity, by a significant transverse momentum
imbalance, and by a pair $e^\pm \mu^\mp$ or $e^\pm \mu^\pm$.

Significant irreducible backgrounds can arise from QCD and electroweak
(EW) processes which lead to the same final state
\begin{displaymath}
p + p \to j  j  e^\pm \mu^\mp \nu \nu \; , \; j  j  e^\pm \mu^\pm \nu \nu \; ,
\end{displaymath}
where the jets arise from a gluon or light quark production.  They
include ``resonant'' processes with the production and subsequent
leptonic decay of $W^\pm W^\mp$ or $W^\pm W^\pm$ pairs (on- or
off--shell) accompanied by jets, and ``non-resonant'' processes
containing only one or no $W$'s in the s--channel. Furthermore for
different sign final leptons, a large QCD background is expected from
the production and subsequent decay of top quark pairs together with
0--2 jets.

The six--particle amplitudes for the signal and irreducible
backgrounds are simulated at the parton level with full tree level
matrix elements.  The SM amplitudes are generated using Madgraph
\cite{mad} in the framework of Helas \cite{helas} routines.  The
anomalous contributions arising from the effective interactions
(\ref{eq:dclin}) and (\ref{eq:dcnolin}) are implemented as subroutines
and included accordingly. We consistently took into account the effect
of all interferences between the anomalous and the SM amplitudes, and
did not use the narrow--width approximation for the vector boson
propagators.  For the treatment of the finite--width effects in
massive vector--boson propagators we use a modified version of the
complex mass scheme \cite{CMS} in which we globally replace
vector-boson masses $m_V^2$ with $m_V^2-im_V \Gamma_V$ without
changing the real value of $\sin^2\theta_W$ \cite{CMS2,dieternew}.
This procedure respects electromagnetic gauge invariance.  We have
also performed a full simulation of the $t \bar{t}$ background and
evaluated the $t \bar{t}$ plus 1 and 2 jets backgrounds using the
narrow width approximation for the top quark. We took the electroweak
parameters $\alpha_{\rm em} = 1/128.93$, $m_Z = 91.189$ GeV, $m_W =
80.419$ GeV, and $m_{\rm top} = 174.3$ GeV. The weak mixing angle was
obtained imposing the tree level relation $\cos \theta_W = m_W/m_Z$,
which leads to $\sin^2 \theta_W = 0.222$. In our calculations we used
CTEQ5L parton distribution functions \cite{CTEQ5_pdf}.

The general expression for the total cross sections for the processes
considered can be written as
\begin{equation}
\sigma=\sigma_{\rm bck}+ g^2 (\Delta c_0)\sigma_0 + g^2 (\Delta c_1)\sigma_1 +
g^4 (\Delta c_0)^2 \sigma_{00}+ g^4 (\Delta c_1)^2 \sigma_{11}+ 
g^4 (\Delta c_0) (\Delta c_1) \sigma_{01} \; ,
\label{eq:sigmatot}
\end{equation}
where $\Delta c_i$ for the case of linear [non-linear] realization of
the gauge symmetry are defined in Eq.\ (\ref{eq:dclin}) [Eq.\
(\ref{eq:dcnolin})].  $\sigma_{\rm bck}$ contains the contributions from all
the backgrounds described above while $\sigma_0$ and $\sigma_1$
contain the interference between SM and anomalous amplitudes. For the
case of a linear realization of the gauge symmetry they contain the
contribution of the light Higgs boson exchange, which is absent in the
non--linear case. In either scenario the anomalous contributions
$\sigma_{0}$, $\sigma_{1}$, $\sigma_{00}$, $\sigma_{01}$, and
$\sigma_{11}$, as well as the EW contribution to $\sigma_{\rm bck}$ in the
absence of a light Higgs boson, do not respect the unitarity of the
partial--wave amplitudes $(a_\ell^I)$ at large subprocess
center--of--mass energies $M_{WW}$ \cite{boos}.  For higher $WW$
invariant masses, rescattering effects are important to unitarize the
amplitudes.  Taking into account this fact, we conservatively we
impose in these cases the cut $M_{WW}<$ 1.25 TeV, which guarantees
that the unitarity constraints are always satisfied.  This requirement
corresponds to a sharp--cutoff unitarization \cite{unit2}.

An important feature of the WBF signal is the absence of color
exchange between the final state quarks, which leads to a depletion of
gluon emission in the region between the two tagging jets. Thus one
can enhance the signal to background ratio by vetoing additional soft
jet activity in the central region \cite{veto}. Certainly, a central
jet veto is ineffective against the EW backgrounds which possess the
same color structure as the signal.  For the QCD backgrounds, however,
there is color exchange in the $t$--channel and consequently a more
abundant production of soft jets, with $p_T>20$~GeV, in the central
region \cite{CZ}. The probability of an event to survive such a
central jet veto has been analyzed for various processes in
Ref.~\cite{rainth}, from which we take the veto survival probabilities
0.8 (0.3) for electroweak (QCD) processes. Moreover, at the
high--luminosity run of the LHC there will be more than one
interaction per bunch crossing, consequently there is a probability of
detecting an extra jet in the gap region due to pile--up.  In
Ref.~\cite{atlas} it was estimated that due to pile--up the jet--veto
efficiency for a threshold cut of $p_T=20$ GeV is 0.75.  Taking into
account these two effects we obtain that the veto survival
probabilities are
\begin{equation}
P_{\rm surv}^{\rm EW}= 0.8 \times 0.75=0.6 \;\;\;\;\;\; , \;\;\;\;\;\;
P_{\rm surv}^{\rm QCD}=0.3 \times 0.75=0.225 \;\; .
\label{eq:psurv}
\end{equation}

Constraining quartic gauge boson couplings in the WBF processes $pp
\to jj e \mu \nu \nu$ is essentially a counting experiment since there
is no resonance in the $WW$ invariant mass distribution. The
sensitivity of the search is thus determined by the precision with
which the background rate in the search region can be predicted. In
order to access the size of these uncertainties we have employed four
different choices of the renormalization and factorization scales
which we denote by:
\begin{itemize}
\item [{\bf C1}] $\mu_F^0 = \mu_R^0=\sqrt{(p_{Tj_1}^2+p_{Tj_2}^2)/2}\;$;
\item [{\bf C2}] $\mu_R^0 = \sqrt{(p_{Tj_1}^2+p_{Tj_2}^2)/2}$ and
  $\mu_F^0 = \sqrt{\hat s}$ where $\hat s$ is the
  squared parton center--of--mass energy;
\item [{\bf C3}] $\mu_R^0 = \mu^0_F =\sqrt{p_{Tj1}~ p_{Tj2}}$;
\item [{\bf C4}]  $\alpha_s^2(\mu_R^0) =
  \alpha_s(p_{Tj_1})~\alpha_s(p_{Tj_2})$ and $\mu_F^0 =
  \hbox{min}(p_{Tj1},~ p_{Tj2})$.
\end{itemize}

Finally, we simulate experimental resolutions by smearing the energies
(but not directions) of all final state partons with a Gaussian error
given by $\Delta E/E = 0.5/\sqrt{E} \oplus 0.02$ if $|\eta_j| \leq3$
and $\Delta E/E = 1/\sqrt{E} \oplus 0.07$ if $|\eta_j| >3$ ($E$ in
GeV), while for charged leptons we used a resolution $\Delta E/E =
0.1/\sqrt{E} \oplus 0.01$.  We considered the jet tagging efficiency
to be $0.75 \times 0.75=0.56$ while the lepton detection efficiency is
taken to be $0.9 \times 0.9 =0.81$.

\section{Signal and  Background Properties}
\label{sec:sigback}

\subsection{Basic cuts}

We initially impose the following jet acceptance cuts
\begin{equation}
 p_T^j > 20 \hbox{ GeV} \;\;\;\;\;\;\; , \;\;\;\; | \eta_j | < 4.9 \;\; ,
\label{cuts1}
\end{equation}
in order to have a well defined tagging jets.  We also demand lepton
acceptance and isolation cuts
\begin{eqnarray}
|\eta_\ell|\leq 2.5 \;\;\; &,& \;\;\;
 \eta_{\rm min}^j < \eta_\ell < \eta_{\rm max}^j 
\nonumber
\\
\Delta R_{\ell j}\geq 0.4 \;\;\; &,& \;\;\;
\Delta R_{\ell \ell}\geq 0.4 \;\; 
\label{basiccuts}
\\
p_{T}^\ell \geq p_T^{\rm min}   &,&  
\nonumber
\end{eqnarray}
where $\eta_{\rm min \;(max)}$ is the minimum (maximum) rapidity of the
 tagging jets
and $p_T^{\rm min} = 100$ (30) GeV for opposite (equal) charge leptons.  Since
the signal events contains undetectable neutrinos that carry some transverse
energy from the event, we also require a missing transverse momentum
\begin{equation}
  p^T_{\text{missing}} \geq 30 \hbox{ GeV} \; .
\label{ptmis}
\end{equation}
The tagging jets are usually well separated in rapidity in the signal,
therefore we demand the existence of a rapidity gap between them
\begin{equation}
| \eta_{j1} - \eta_{j2} | > 3.8 \;\;\; , 
\;\;\;   \eta_{j1} \cdot \eta_{j2} < 0 \;\; .
\label{cut:rap}
\end{equation}
%

\subsection{Additional cuts for $pp \to jj e^\pm \mu^\mp\nu\nu  $}

The production of opposite sign leptons exhibits a very large background 
due to the production of $t \bar{t}$ pairs in association with $0,1,2$ jets.
In the $t \bar{t}$ process the $b$-quarks produced in the $t$ decays
are identified as the tagging jets. We denote by $t \bar{t}j$ and 
$t \bar{t}jj$ backgrounds those events where the additional 
jet(s) is (are) identified as the tagging jet(s) while one of the 
$b$-jets  in $t \bar{t}j$  and both in $t \bar{t}jj$ are soft and central.
$t \bar{t}j$  and $t \bar{t}jj$ events where one (or
two) of the additional jet are not identified as the tagging ones 
contribute to the QCD radiation of the corresponding  
$t \bar{t}$ and  $t \bar{t}j$ background and their effect is included in the 
gap survival probabilities~\cite{dieter}.  

\begin{figure}[!t]
       \centerline{\psfig{figure=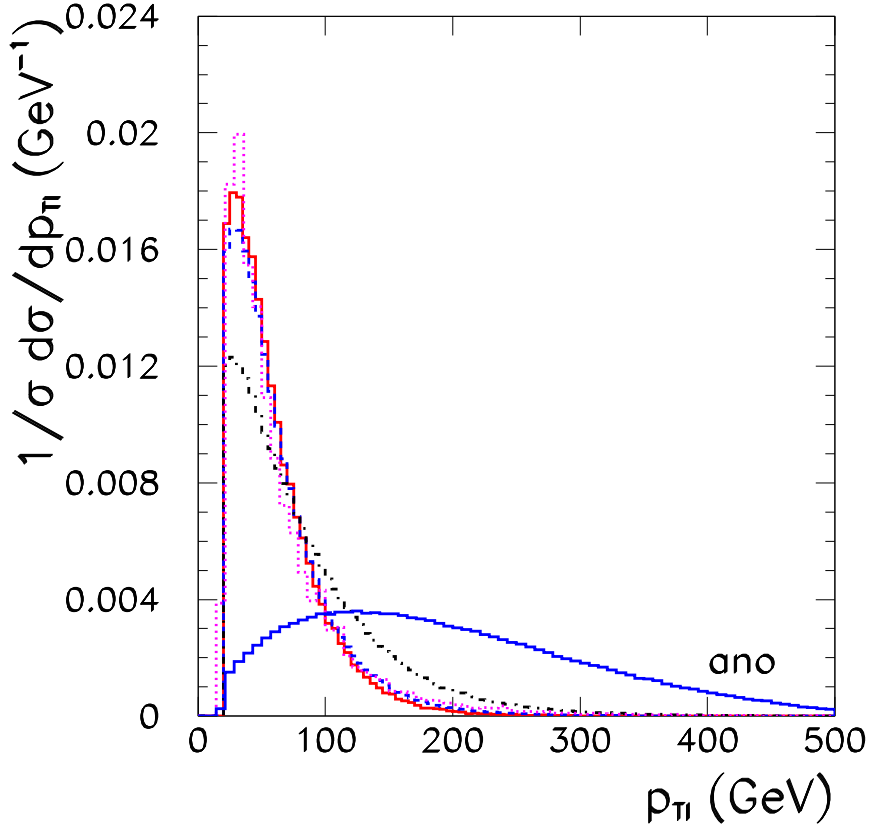,width=5in}}
       \caption{Normalized distribution of the transverse momentum of
         the charge leptons for $t\bar{t}$ (red solid line), $t
         \bar{t}j$ (blue dashed line), $t \bar{t}jj$ (black dot-dashed
         line), SM irreducible production (magenta dotted line), and
         anomalous $W^+ W^-$ contribution $\sigma_{00}$ (blue solid
         line marked ``ano''). We assumed $m_h=120$ GeV and applied
         cuts (\ref{cuts1})--(\ref{cut:rap}) but for with a relaxed cut
         $p_T^\ell>20$ GeV.}
\label{fig:ptl_pm}
\end{figure}

\begin{figure}[!t]
  \centerline{\psfig{figure=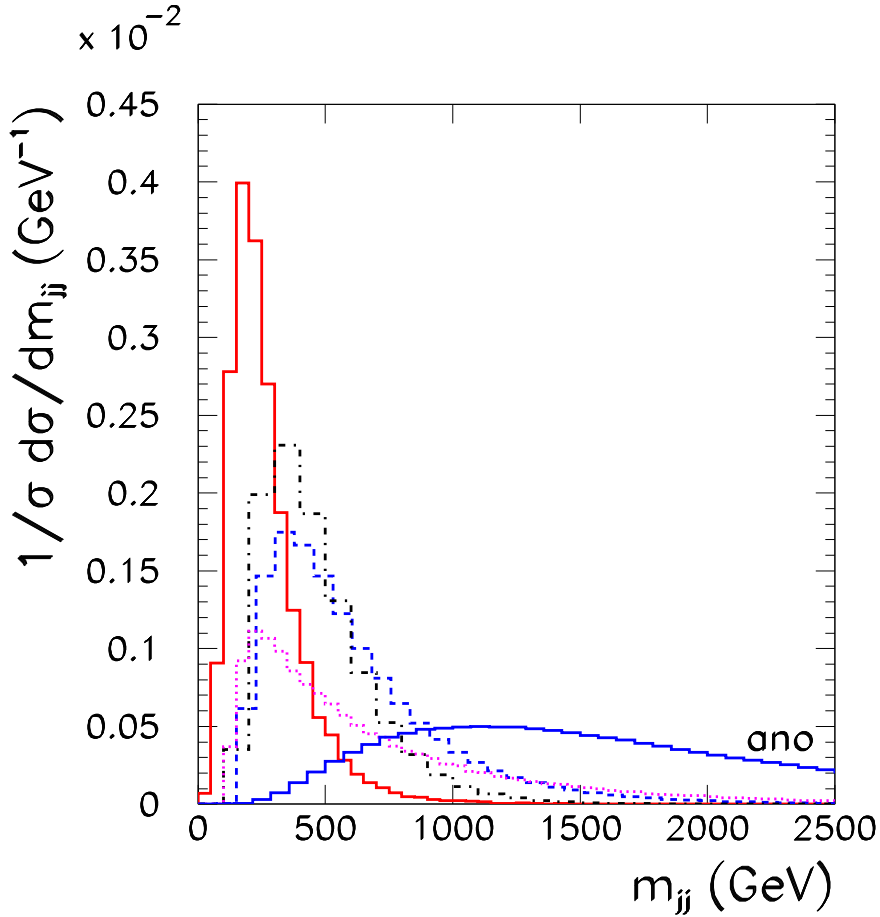,width=5in}}
  \caption{Normalized jet-jet invariant mass distribution for
    $t\bar{t}$ (red solid line), $t \bar{t}j$ (blue dashed line), $t
    \bar{t}jj$ (black dot-dashed line), SM irreducible production
    (magenta dotted line), and anomalous $W^+ W^-$ contribution
    $\sigma_{00}$ (blue solid line marked ``ano''). We assumed
    $m_h=120$ GeV and applied cuts (\ref{cuts1})--(\ref{cut:rap}).}
\label{fig:mjj}
\end{figure}

The relevance of the tighter cut on the transverse  lepton momentum to 
suppress the different backgrounds in  $pp \to jj e^\pm \mu^\mp\nu\nu$ is
illustrated in  Fig.~\ref{fig:ptl_pm}.
In order to further reduce these backgrounds we make use of the fact
that QCD processes typically occur at smaller invariant masses of
tagging jets compared to EW processes. This is illustrated in
Fig.~\ref{fig:mjj} where we show the normalized invariant mass
distribution of the tagging jets for the different backgrounds and the
anomalous contribution $\sigma_{00}$ for $pp\rightarrow jj e^+ \mu^-
\nu\nu$.  Consequently, in order to further suppress the backgrounds
we also require a large invariant mass of the tagging jets
\begin{equation}
       M_{jj}\geq 1000 \hbox{ GeV}  \; ,
\label{eq:cutsqcd}
\end{equation}
which mainly reduces the $t\bar{t}$ events  but still leaves a 
large background from $t\bar{t}j$ and $t\bar{t}jj$ production.
These events can be very efficiently suppressed by vetoing additional 
soft jet activity in the central region. Consequently, we impose
that the event does not contain additional jets with transverse momentum 
larger than 20 GeV in between the tagging ones, 
\begin{equation}
  p_T^j < 20~\hbox{ GeV} \;\;\;\; \hbox{if} \;\;\;\; 
 \eta_{\rm min}^j < \eta_j < \eta_{\rm max}^j \; .
\label{c:veto}
\end{equation}

\begin{figure}[!t]
       \centerline{\psfig{figure=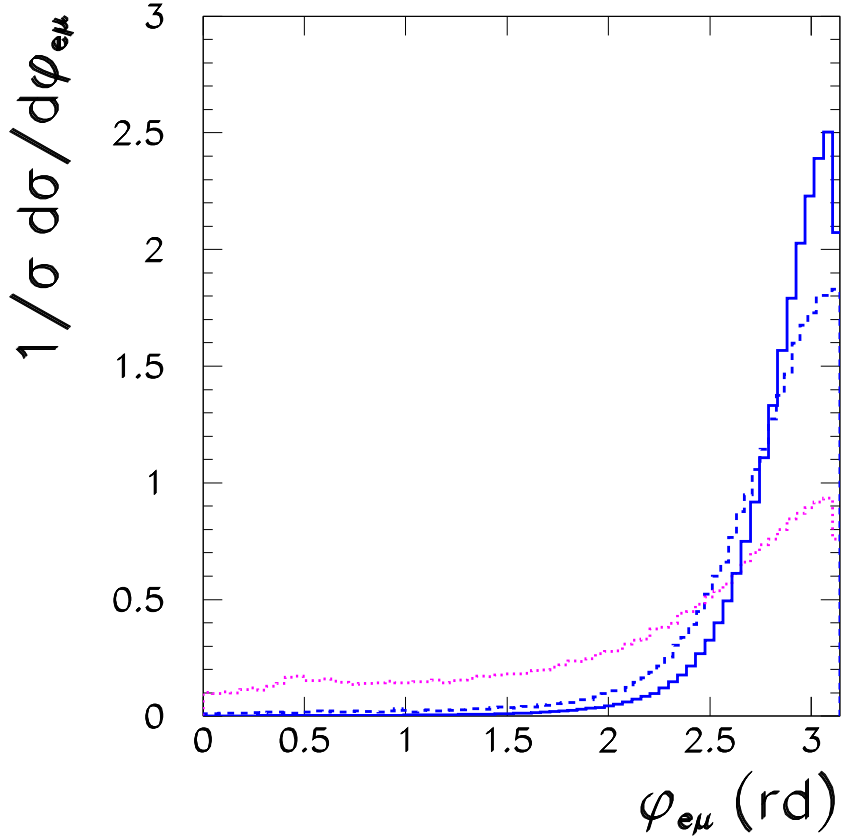,width=5in}}
       \caption{Normalized distribution of the $e\mu$ azimuthal angle
         difference for $t \bar{t}j$ (dashed line), irreducible background 
         (dotted line), and anomalous $W^+ W^-$ production (solid line). We
         assumed $m_h=120$ GeV and applied cuts (\ref{cuts1})--(\ref{c:veto}).}
\label{fig:phi}
\end{figure}

Additionally we notice that the azimuthal angular distribution of the
charged leptons relative to each other in the SM is different than in
the anomalous contributions. The $e^\pm\mu^\mp$ pairs from the decay
of the $W$ pairs produced via the effective interactions
(\ref{eq:dclin}) and (\ref{eq:dcnolin}) are preferentially emitted in
opposite direction from each other. This is shown in
Fig.~\ref{fig:phi} where we plot the normalized distribution of the
azimuthal angle between the electron and the muon.  Thus we impose
also the additional cut
\begin{equation}
\varphi_{e\mu} > 2.25\; {\rm rd}.
\label{c:phi}
\end{equation}

\begin{figure}[!t]
       \centerline{\psfig{figure=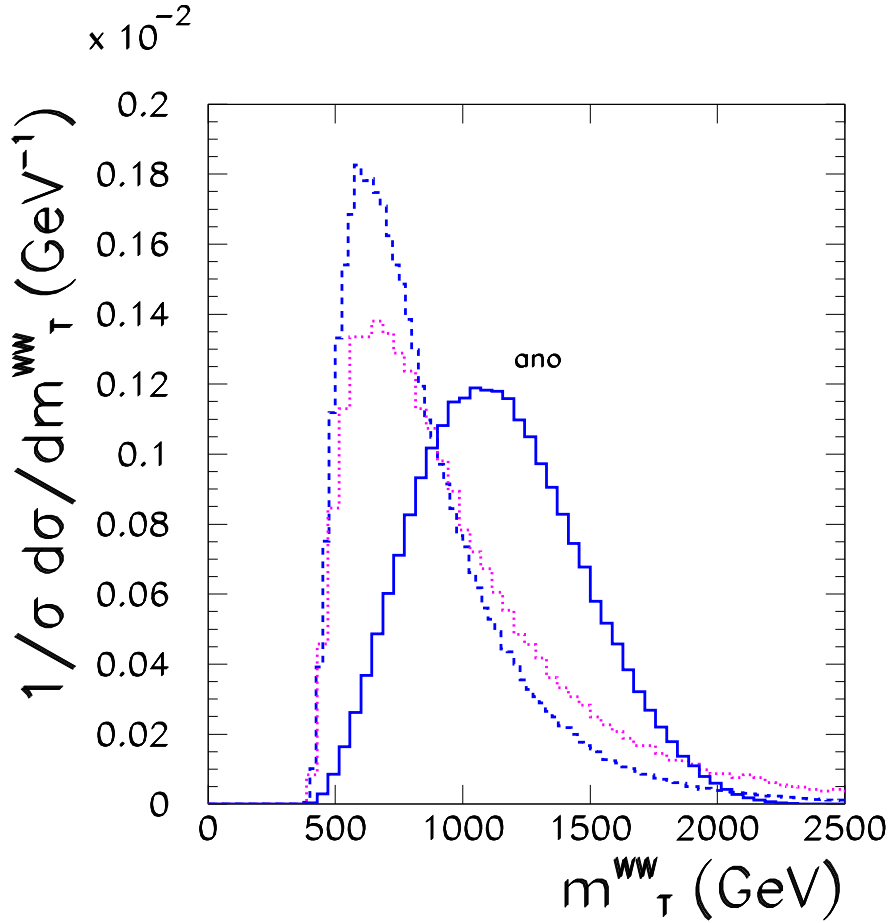,width=5in}}
       \caption{Normalized distribution of $M_T^{WW}$ 
         for $t \bar{t}j$ (dashed line), irreducible background 
         (dotted line), and anomalous $W^+ W^-$ production (solid line). We
         assumed $m_h=120$ GeV and applied cuts (\ref{cuts1})--(\ref{c:phi}).}
\label{fig:mtww}
\end{figure}

Finally we make use of the fact that the anomalous contributions
arising from the effective interactions (\ref{eq:dclin}) and
(\ref{eq:dcnolin}) lead to a growth of the cross section for large
$WW$ invariant masses; see Fig.~\ref{fig:mtww}.
Consequently we define the signal region 
\begin{equation}
  M_T^{WW} \ge 800 \hbox{ GeV} \; ,
\label{c:mtww}
\end{equation}
where we define the transverse invariant mass as
\begin{equation}
  M_T^{WW} = \left( \sqrt{(p_T^{e\mu})^2 + m^2_{e\mu}}   
              + \sqrt{\sla{p_T}^2 + m^2_{e \mu} }\right)^2
 - (\vec{p_T^{e\mu}} + \vec{\sla{p_T}}  )^2
\label{eq:mtww}
\end{equation}
where $\vec{\sla{p_T}}$ is the missing transverse momentum vector,
$\vec{p_T^{e\mu}}$ is the transverse momentum of the pair $e$-$\mu$
and $m_{e \mu}$ is the $e\mu$ invariant mass.

In Table \ref{tab:cutsos}, we illustrate the effect of the above cuts
for $pp \to jj e^\pm \mu^\mp\nu\nu $.  In the lines marked IRED$+-$ we
take into account the full scattering amplitude for the irreducible
backgrounds.
We separate the electroweak and QCD
part of these backgrounds in order to show the effect of the veto 
survival probabilities.
As illustration of the signal loss due to the imposed cuts, we also include 
the cross section for the anomalous term $\sigma_{00}$.
From this table, we can see that 
the largest background is the $t \bar{t} + n$
jets production by three order of magnitude when we apply only the acceptance
and tagging cuts. However, after cuts the dominant backgrounds 
are $t \bar{t} j$ and EW irreducible processes.

\begin{table}[ht]
\begin{center}
\begin{tabular}{|l|l||l|l|l|l|l||}
\hline
background/cut  
&  (\ref{cuts1})--(\ref {cut:rap}) [$20$ GeV] 
&  (\ref{cuts1})--(\ref {cut:rap})  
& (\ref{cuts1})--(\ref{eq:cutsqcd})  
&  (\ref{cuts1})--(\ref {c:veto}) 
&  (\ref{cuts1})--(\ref {c:veto}) $\times P_{\rm surv}$ 
& (\ref{cuts1})--(\ref{c:phi}) 
$\times P_{\rm surv}$ 
\\
\hline \hline
IRED$+-$ (QCD) & 20.0  & 1.12 & 0.26  &  0.26 & 0.058 & 0.035  
\\
\hline
IRED$+-$ (EW) & 4.4 &  0.30  & 0.24 & 0.24 &  0.14& 0.089  
\\
\hline
$t \bar{t}$   & 217. & 6.96 & 0.0306 & 0.0306 & 0.0069 & 0.0068
\\
\hline
$t \bar{t} j$ 
& 1860.  & 73.8 & 8.88 & 0.776 & 0.175  & 0.158
\\
\hline
$t \bar{t} jj$ & 682. & 77.2 & 2.21  & 0.0140 & 0.0032 & 0.0031
\\
\hline \hline
Anomalous $\sigma_{00}$ & 2710& 1710 & 1310 & 1310 & 786 & 758
\\
\hline \hline 
\end{tabular}
\end{center}
\caption{Cross sections in fb to illustrate the effect of cuts for the 
  $pp \to j j e^\pm \mu^\mp\nu\nu  $ production.
  The column marked as (\ref{cuts1})--(\ref {cut:rap}) [$20$ GeV] 
  shows the total
  cross sections after applying out basic cuts with a relaxed
  $p_T^{\rm min} = 20$  GeV. In computing the SM cross section  the 
  contribution from a 
  light Higgs boson with $m_h=120$ GeV is included.
  The cross sections (given in fb) do not include the forward jet 
  and lepton 
  detection efficiencies and they are obtained for choice {\bf C1}
  of the renormalization and factorization scales.}
\label{tab:cutsos}
\end{table}

\subsection{Cuts for $pp \to  jj e^\pm \mu^\pm \nu\nu$}

In the first column in Table \ref{tab:cutsss} we give the cross
sections for $pp \to jj e^\pm \mu^\pm \nu\nu $ after the basic cuts
(\ref{cuts1})--(\ref{cut:rap}). As we can see, the only important source
of background events is the SM electroweak processes contributing to
the same final state.

\begin{figure}[!t]
\centerline{\psfig{figure=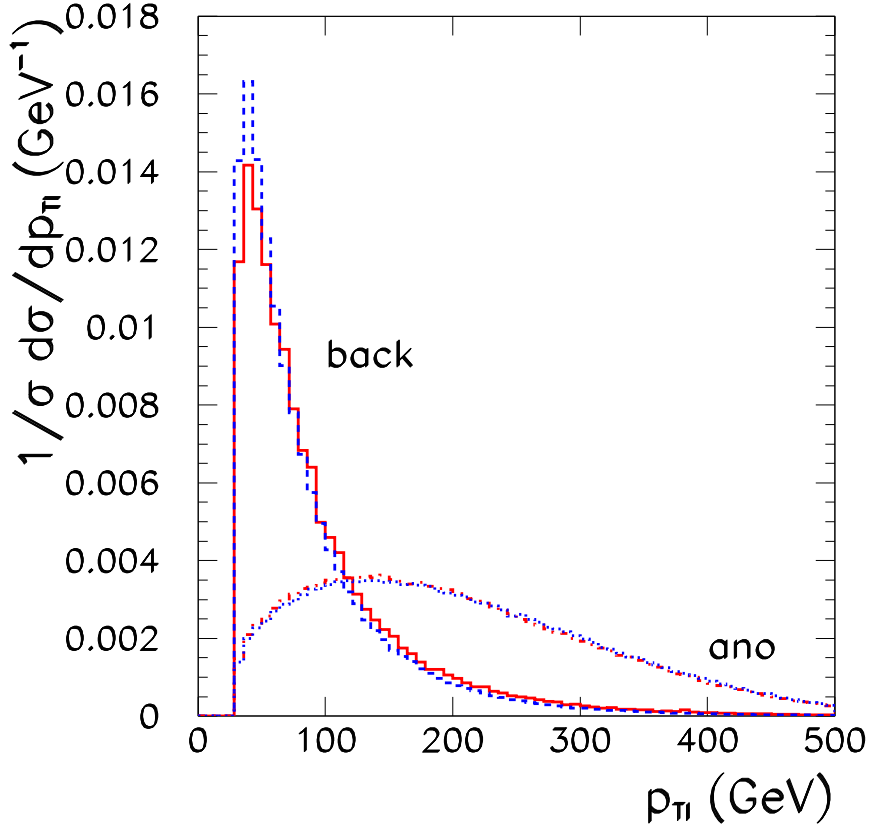,width=5in}} \caption{Normalized
lepton transverse momentum distribution for IRED$++$ background (solid
line), IRED$--$ background (dashed line), $W^+W^+$ $\sigma_{00}$
(dot-dashed line), and $W^- W^-$ $\sigma_{00}$ (dotted line). We
assumed $m_h=120$ GeV and applied cuts (\ref{cuts1})--(\ref{cut:rap}).}
\label{fig:ptl}
\end{figure}

Further enhancement of the signal from 
the anomalous contributions can be obtained by studying the transverse
momentum of the produced leptons as demonstrated in 
Fig.~\ref{fig:ptl} 
which shows the lepton transverse momentum distribution
for the anomalous contribution $\sigma_{00}$ and for the SM background. As we
can see, the background is peaked toward small lepton transverse momenta while
the anomalous contributions leads to the production of leptons with a higher
transverse momentum.
Consequently, for processes leading to final state leptons with the same 
charge, we define our signal region by tightening the $p_T^\ell$ cut
\begin{equation}
 p_T^\ell > 100~ \hbox{GeV.} 
\label{cut:wpwp}
\end{equation}
\begin{table}[ht]
\begin{center}
\begin{tabular}{|l||l|l|l||}
\hline
background/cut  & (\ref{cuts1})--(\ref{cut:rap}) 
&(\ref{cuts1})--(\ref{cut:rap}) $+$ (\ref{cut:wpwp}) 
&[(\ref{cuts1})--(\ref{cut:rap}) $+$ (\ref{cut:wpwp})]
$\times P_{\rm surv}$\\
\hline \hline
IRED$++$ (QCD) & 0.07  & 0.004& 0.0009
\\
\hline 
IRED$++$ (EW) & 1.11  & 0.105 & 0.063
\\
\hline \hline
Anomalous $\sigma_{00}$ & 2250. &  1470.& 880.
\\
\hline \hline
IRED$--$ (QCD)    & 0.025 & 0.001& 0.0002
\\
\hline
IRED$--$ (EW)   & 0.365  & 0.046 & 0.028
\\
\hline \hline
Anomalous $\sigma_{00}$ & 536. & 334. & 200.
\\
\hline \hline
\end{tabular}
\end{center}
\caption{Effect of cuts for $pp \to j j e^\pm \mu^\pm\nu\nu  $ 
production.
The cross sections (in fb) do not include the forward jet and lepton 
detection efficiencies 
and they are obtained for choice {\bf C1}
of the renormalization and factorization scales.}
\label{tab:cutsss}
\end{table}

We present in Table \ref{tab:sig} our final results for the
coefficients $\sigma_{\rm bck}, \sigma_i, \sigma_{i,j}$ of
Eq.~(\ref{eq:sigmatot}) after cuts (\ref{cuts1})--(\ref{c:phi}) for
$pp\rightarrow jj e^\pm \mu^\mp \nu\nu$ and cuts
(\ref{cuts1})--(\ref{cut:rap}) $+$ (\ref{cut:wpwp}) for $pp\rightarrow jj
e^\pm \mu^\pm \nu\nu$.  The results include the effect of the veto
survival probabilities as well as the forward jet and lepton detection
efficiencies.  They were obtained for choice {\bf C1} of the
renormalization and factorization scales.

\begin{table}[ht]
\begin{center}
\begin{tabular}{|l|l||l|l|l|l|l|l||}
\hline
scenario & channel & $\sigma_{\rm bck}$ &
$\sigma_0$ &$\sigma_1$ &$\sigma_{00}$ &$\sigma_{11}$ 
&$\sigma_{01}$ \\ 
\hline\hline 
& $pp \to e^\pm \mu^\mp \nu\nu j j$ 
& 0.067 & ---    & ---   & 300 & 655 & 822
\\ \hline 
$m_h=120$ GeV &
$pp \to e^+ \mu^+ \nu\nu j j$  
& 0.029 &  $-0.46$ & $-0.20$ & 400 &  94 & 380 \\ \hline
& $pp \to e^- \mu^- \nu\nu j j$   
& 0.045 &  $-0.11$ & $-0.04$ &  91 &  21 &  87 
\\ \hline\hline
& $pp \to e^\pm \mu^\mp \nu\nu j j$  & 0.07 &  1.3  & 2.1 
& 300 & 655 & 822\\  \hline
No light Higgs boson  
& $pp \to e^+ \mu^+ \nu\nu j j$   
& 0.017&  $-4.9$ &$-2.3$ & 400 &  94 & 380 \\ \hline
& $pp \to e^- \mu^- \nu\nu j j$   & 0.017&  $-1.2$ & $-0.54$ &  91 &  21 &  87 
\\ \hline\hline
\end{tabular}
\end{center}
\caption{Cross sections (in fb) for the different terms in 
Eq.~(\ref{eq:sigmatot}). 
The results include the effect of the 
veto survival probabilities and the forward jet and lepton 
detection efficiencies. 
They are obtained for choice {\bf C1}
of the renormalization and factorization scales.}
\label{tab:sig}
\end{table}

\subsection{Estimating the Backgrounds}

As mentioned above, constraining the quartic gauge boson couplings is
essentially a counting experiment and the sensitivity of the search is thus
determined by the precision with which the background rate in the search
region can be predicted. Since the signal selection is demanding, including
double forward jet tagging and central jet vetoing techniques whose acceptance
cannot be calculated with sufficient precision in perturbative QCD, the
theoretically predicted background can vary up to a large factor.  Though the
QCD corrections to the irreducible EW processes seem to be
modest~\cite{dieternew} the same is not guarantee for the QCD backgrounds.
Here the situation is analogous to the Higgs boson production in WBF
  where the backgrounds must be also estimated from data \cite{estback}.

\begin{figure}[!t]
\centerline{\psfig{figure=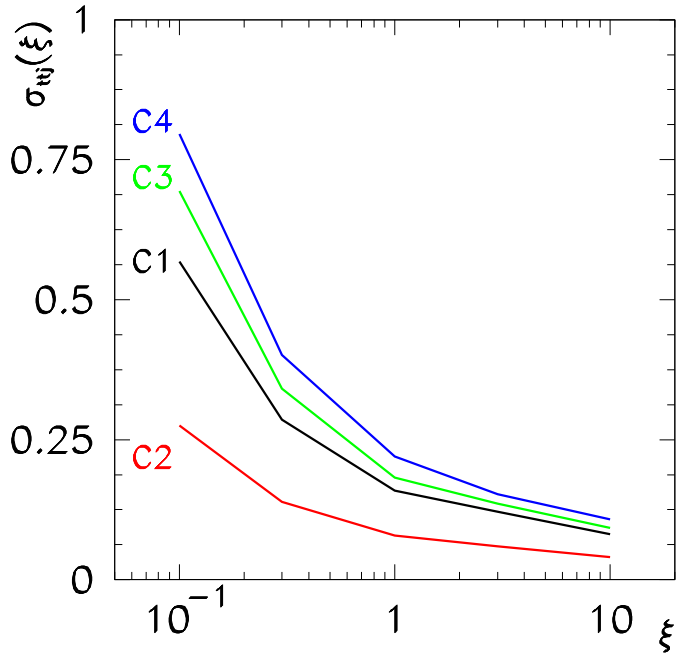,width=4in}} 
\caption{Cross section for $t\bar{t}j$ after cuts 
(\ref{cuts1})--(\ref{c:phi}) as a function 
$\xi$, where $\mu_R=\xi\mu_R^0$ for the four choices
 choices of the factorization and renormalization scale defined
in the text.}
\label{fig:ttjcross}
\end{figure}

We demonstrate the large QCD uncertainties in Fig.~\ref{fig:ttjcross}
where we plot the value of the $t\bar{t}j$ cross section after cuts
(\ref{cuts1})--(\ref{c:phi}) for different choices of the factorization
and renormalization scale. Moreover, we should also keep in
  mind that the narrow width approximation used by us has a
  discrepancy with respect to the full LO amplitude calculation of the
  order of 10--20\% \cite{nwafull}. The obvious conclusion is that in
order to obtain a meaningful estimate of the sensitivity the
background levels need to be determined directly from LHC data.
Fortunately, a sizable sample of $ jj e^\pm \mu^\mp\nu\nu$ and $ jj
e^\pm \mu^\pm \nu\nu$ events will be available if some of the cuts are
relaxed. In this way the background normalization error can be reduced
by considering a larger phase space region as a calibration region.
The background expected in the signal region is then obtained by
extrapolation of the measured events in the calibration region to the
signal region. This procedure introduces also an uncertainty, which we
denote as QCD--extrapolation uncertainty, due to the extrapolation to
the signal region. However, as we will show, these uncertainties are
smaller than the overall normalization uncertainty.

Using the results in Fig.~\ref{fig:mtww} we see that we can define the
calibration region used to estimate the background for $ jj e^\pm
\mu^\mp\nu\nu$ as the one complying with cuts
(\ref{cuts1})--(\ref{c:phi}) and 
\begin{equation}
 M_T^{WW} \le 800 \hbox{ GeV.}
\label{eq:contregion}
\end{equation}
Equivalently from the results in Fig.~\ref{fig:ptl} we find that one
can define the calibration region used to estimate the background for
$ jj e^\pm \mu^\pm\nu\nu$ as the one within cuts
(\ref{cuts1})--(\ref{cut:rap})
\begin{equation}
30 < p_T^\ell <100~ \hbox{GeV.} 
\end{equation}

As a measure of theoretical uncertainty associated with the
extrapolation from the calibration to the signal regions, we study the
ratio of the cross sections in the signal region and the calibration
region as a function of $\xi$, the scale factor for the four different
renormalization scale choices $\mu_R = \xi\mu_R^0$ listed above. In
this way we define for the final state exhibiting opposite charge
leptons ($ jj e^\pm \mu^\mp\nu\nu$)
\begin{equation}
  R_{\rm os} = 
\frac{\displaystyle\sigma_{\rm bck}(M_T^{WW}> 800 \hbox{ GeV} )}
     {\displaystyle\sigma_{\rm bck}(M_T^{WW}< 800 \hbox{ GeV})} \; ,
\end{equation}
where in the evaluation of these ratios we have added the electroweak
and QCD contributions from all background sources taking into account
the corresponding veto survival probabilities.  
On the other hand, for
the final state $ jj e^\pm \mu^\pm\nu\nu$ that exhibits same charge
leptons we define
\begin{equation}
   R_{\rm ss} = \frac{\displaystyle \sigma_{\rm bck}(p_T^\ell>100~{\rm GeV})}
                {\displaystyle\sigma_{\rm bck}(30<p_T^\ell<100~{\rm GeV})}
 \; ,
\end{equation}
where we have added the contributions from both signs.

\begin{figure}[!t]
\centerline{\psfig{figure=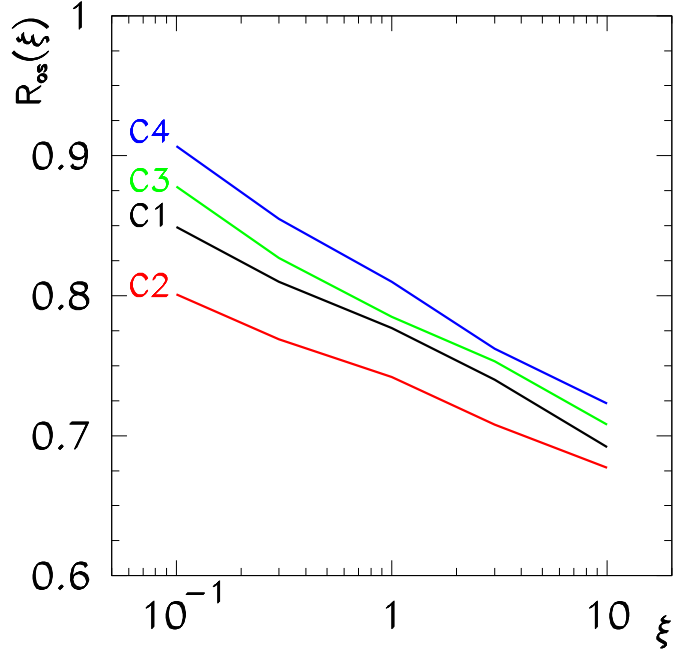,width=4in}} 
\caption{ The ratio $R_{\rm os}$ is shown as a function of $\xi$,
  where $\mu_R=\xi\mu_R^0$ for the four choices of renormalization and
  factorization scales given in the text.}
\label{fig:scale}
\end{figure}

We depict in Fig.~\ref{fig:scale} the $\xi$ dependence of $R_{\rm os}$
which shows that the extrapolation uncertainty is at a tolerable level
($\simeq 15$\%) being much smaller than the normalization uncertainty.
The corresponding extrapolation uncertainty for the processes with
same sign leptons is smaller by a factor of 2 because of the QCD
background is small.

Altogether the total expected uncertainty in the estimated number of
background events has two sources: the {\sl theoretical} uncertainty
associated to the extrapolations from the calibration region
($\delta_{\text{bck,th}}$) and the {\sl statistical} error associated
to the determination of the background cross section in the
calibration region ($\delta_{\text{bck,stat}}$). This last one is
slightly different for the case of light or no light Higgs boson
because of the slightly different number of events from the SM
irreducible background.  Assuming an integrated luminosity of 100
fb$^{-1}$ we find
\begin{eqnarray}
\delta_{\text{bck,th,os}}=15  \%\;\;,\;\;\;  
&  
\delta^{\rm lin}
_{\text{bck,stat,os}}=34\% \;\;,\;\;\;  
& \delta^{\rm no-lin}_{\text{bck,stat,os}}=31\% \;\;, \\
\delta_{\text{bck,th,ss}}=7.5 \% \;\;,\;\;\;   
&  \delta^{\rm lin}_{\text{bck,stat,ss}}=22\% \;\;,\;\;\;  & 
\delta^{\rm no-lin}_{\text{bck,stat,ss}}=21\% \;\;,
\end{eqnarray}
where we denoted by the superscript ``lin'' (``non-lin'') the case
with (without) a light Higgs boson. In addition to these
  uncertainties considered here there are also experimental systematic
  uncertainties, which are sizable for the Higgs boson searches
  \cite{estback}, however they do require a full detector simulation
  which is beyond the scope of this work.

%

\section{Results and Discussion}
\label{sec:results}

In order to obtain the attainable sensitivity to deviations of the SM
predictions of the quartic gauge boson couplings we assumed an
integrated luminosity of $100$ fb$^{-1}$ and that the observed number
of events in the different scenarios is compatible with the background
expectations for the choice {\bf C1} of the renormalization and
factorization scales both in the signal  
($N^{\text{S}}_{i, \text{data}}$) and in the calibration ($
N^\text{C}_{i, \text{data}}$) regions, {\em i.e.}
\begin{equation}
N^{\text{S}}_{i, \rm data}= N^{\text{S}}_{i, \rm bck,C1} 
\;\;\;\;\;\hbox{and}\;\;\;\;\; 
N^{\text{C}}_{i, \rm data}=N^{\text{C}}_{i, \rm bck,C1} \;\; ,
\label{hypo}
\end{equation}
where we denote the process $pp\rightarrow j j e^\pm \mu^\mp \nu\nu$
by $i=$ os and the sum of $pp\rightarrow j j e^+ \mu^+ \nu\nu $ and
$pp\rightarrow j j e^- \mu^- \nu\nu$ by $i=$ ss.

Deviations from the SM prediction for the four gauge boson vertices
manifest themselves as a difference between the number of observed
events and the number of background events estimated from the
extrapolation of the background measured in the calibration region ($
N_{i, \rm back}^{\text{S}}$), that is,
\begin{equation}
     N^{\text{S}}_{i, \rm data} - N_{i, \rm back}^{\text{S}} \;\; ,
\label{sig:def}
\end{equation}
where $N_{i, \rm back}^{\text{S}}= R_i \, N^{\text{C}}_{i, \rm data}$.
Notice that (\ref{hypo}) implies that we are assuming that no
  departure of the SM predictions has been observed neither in the
  control region nor in the signal one.

The statistical error of the number of anomalous events is
\begin{equation}
   \sigma_{i,\text{stat}}^2 = N_{i, \text{data}}^{\text{S}}
     + (R_i \, N_{i, \text{data}}^\text{C} \delta_{\text{bck,stat}})^2
\label{err:stat}
\end{equation}
where the first term is the statistical error of the measured number of events
in the signal region and the second term is the error in the determination of
the background in the signal region due to the statistical error of the
background measurement in the calibration region, $\delta_{\text{bck,stat}}$.
The extrapolation uncertainty introduces an additional error 
\begin{equation}
   \sigma_{i,\text{th}}=R_i\, N_{i, \text{data}}^\text{C} 
\delta_{\text{bck,th}}   \;\; .
\label{err:th}
\end{equation}
Both errors can be assumed to be Gaussian and we combine then in quadrature.

\begin{figure}[!t]
\centerline{\psfig{figure=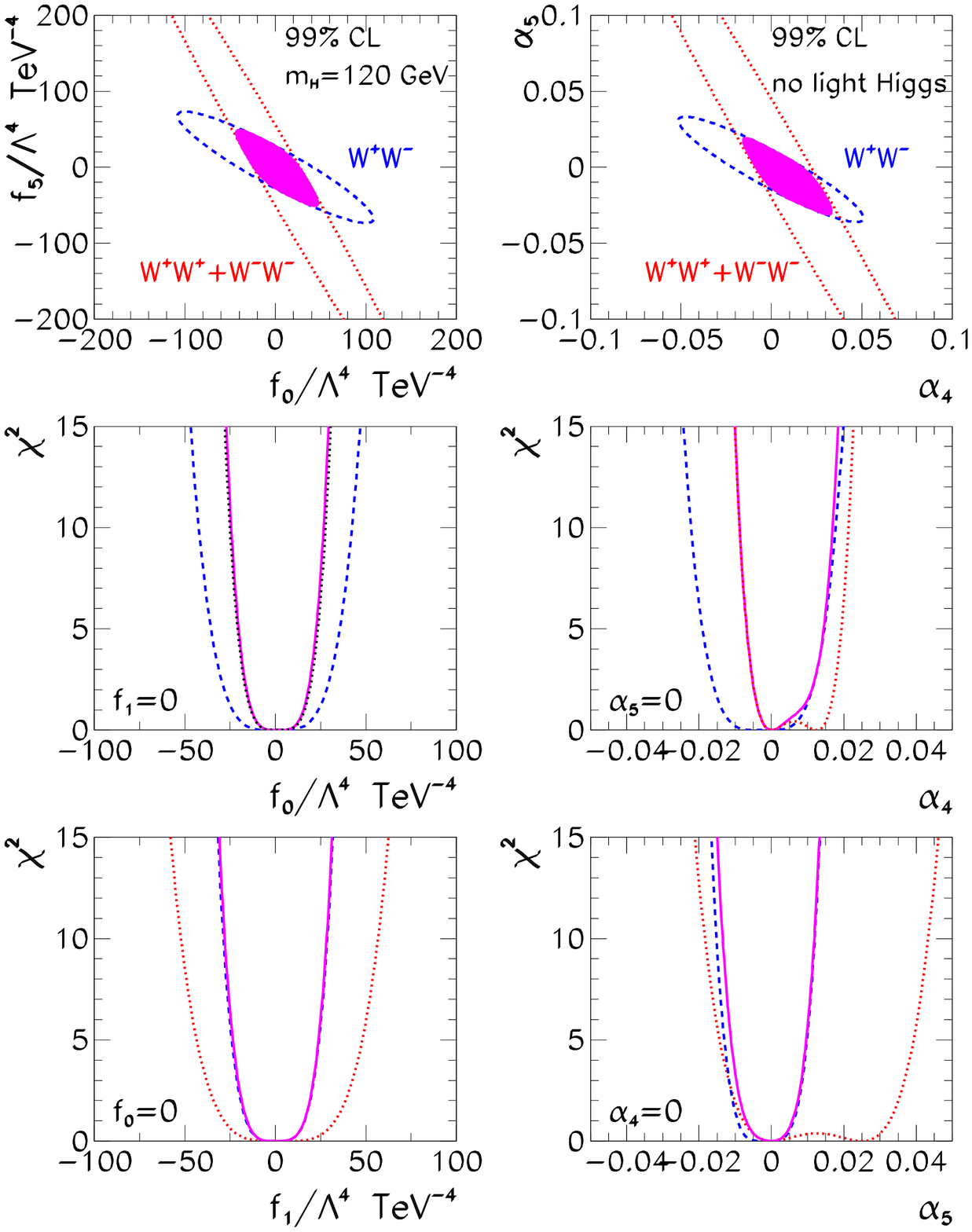,width=0.9\textwidth}} 
\caption{Sensitivity bounds on the coefficients of the anomalous
  quartic gauge boson operators for the case with a light Higgs boson
  (left panels) and with no light Higgs boson (right panels).  The
  upper panels show the 99\% CL allowed regions for the different
  channels.
The lower panels show the dependence of the $\chi^2$
  function for the different channels assuming that only under that
  only one anomalous parameter is non--vanishing.  
 The dashed (dotted) lines corresponds to 
$pp\rightarrow j j e^\pm \mu^\mp \nu\nu$  
($pp\rightarrow j j e^\pm \mu^\pm \nu\nu$) while the combined analysis
is indicated by the filled region and solid lines. }   
\label{fig:contours}
\end{figure}

Given our definition of the signal (\ref{sig:def}), the errors
(\ref{err:stat})--(\ref{err:th}), and the parametrization of the cross
section in Eq.~(\ref{eq:sigmatot}) we can easily obtain the attainable
limits on any combination of quartic anomalous coefficients. We
exhibit in the upper panels of Fig.~\ref{fig:contours} the 99\% CL
exclusion region in the plane $f_0$ versus $f_1$ (left) and $\alpha_4$
versus $\alpha_5$ (right) for each channel $i$ independently, and for
the combination of both (full region). As we can see the same sign
processes present a very strong correlation between both couplings
while the correlation is somewhat smaller for the case of the
processes with opposite sign leptons. As a consequence, the final
allowed regions are rather ``compact'' and meaningful sensitivity
bounds can be derived.

In the lower panels of Fig.~\ref{fig:contours}  we plot the
$\chi^2$ as a function of individual couplings, under the assumption
that only one anomalous parameter is non--vanishing. From this we find
that for the case with a light Higgs boson of $m_h=120$ GeV 
\begin{eqnarray}
   -22 < &\frac{\displaystyle f_0}{\displaystyle\Lambda^4} 
({\rm TeV}^{-4})& <24 \;\; ,
\\
 -25 < &\frac{\displaystyle f_1}{\displaystyle \Lambda^4} 
({\rm TeV}^{-4}) & <25 \;\; ,
\end{eqnarray} 
at 99\% CL. In models without a light Higgs boson we get the following
99\% CL bounds
\begin{eqnarray}
 -7.7\times 10^{-3} < &\alpha_4& <15 \times 10^{-3} \;\; ,
%
\\
 -12\times 10^{-3} < &\alpha_5& <10 \times 10^{-3} \;\; .
\end{eqnarray}

These results represent an improvement of more than one order of
magnitude over the present sensitivity from indirect effects in low
energy observables (\ref{eq:lel}) and (\ref{eq:lenl}).
Notwithstanding, for the case in which a light Higgs boson is found
and the gauge theory is linearly realized, they do not reach the
expected natural order of magnitude $f_i\sim{\cal O}(1)$ for new
physics scale above 1 TeV.  On the other hand, for scenarios without a
light Higgs boson a natural order of magnitude of the anomalous
couplings $\alpha_i$ in a fundamental gauge theory is $ g^2
v^2/\Lambda^2$ \cite{aew}, since the quartic anomalous interactions
can be generated by tree diagrams.  Thus, we might expect that the
size of the $\alpha$'s should be of the order of $M_Z^2/\Lambda^2
\simeq 2 \times 10^{-3}$ which is close to the attainable sensitivity
that we obtain from our analysis.

It is also interesting to notice that the achievable sensitivity at
the LHC is close to the recently derived lower bounds based on the
usual analytical properties associated with causal, unitary theories
\cite{grinstein}.  The lack of observation of an anomalous coupling
$\alpha_4$ and $\alpha_5$ below that bound, would indicate the
breakdown of some of these basic properties of the $S$-matrix.  In
particular, as pointed out in Ref.~\cite{grinstein}, since String
theory is designed to produce $S$-matrix with these properties and
therefore, the experimental verification of those bounds could be use
to falsify string theory.


\vskip1cm

\noindent{\bf Acknowledgements:}
We would like to thank D.~Rainwater and N.~Kauer for illuminating discussions. 
This work was partially supported by Funda\c{c}\~{a}o de Amparo \`a
Pesquisa do Estado de S\~ao Paulo (FAPESP), by Conselho Nacional de
Desenvolvimento Cient\'{\i}fico e Tec\-no\-l\'o\-gi\-co (CNPq).  MCG-G
is supported by National Science Foundation grant PHY-0354776 and by
Spanish Grant FPA-2004-00996.

\vskip1cm


\appendix
\section{Dimension 8 effective operators}\label{app:dim8}
We list here the parity conserving effective Lagrangians leading to pure
quartic couplings between the weak gauge bosons assuming that a Higgs boson
has been discovered, that is, employing the linear representation for the
higher order operators. Denoting by $\Phi$ the Higgs doublet and by $U$ an
arbitrary $SU(2)_L$ transformation, the basic blocks for constructing the
effective Lagrangian and their transformations are:
\begin{eqnarray}
     \Phi \;, \; \hbox{ that transforms as } && \Phi^\prime = U \Phi
\\
     D_\mu \Phi \;, \; \hbox{ that transforms as }  &&
 D^\prime_\mu \Phi^\prime = U D_\mu \Phi
\\
 \hat{W}_{\mu\nu} \equiv \sum_j W^j_{\mu\nu} \frac{\sigma^j}{2}
    \;, \; \hbox{ that transforms as }  &&
 \hat{W}_{\mu\nu}^\prime = U  \hat{W}_{\mu\nu} U^\dagger
\\
B_{\mu\nu} \;, \; \hbox{ that transforms as } && B_{\mu\nu}^\prime =
B_{\mu\nu}
\end{eqnarray}
where $W^i_{\mu\nu}$ is the $SU(2)_L$ field strength and $B_{\mu\nu}$
is the $U(1)_Y$ one. The covariant derivative is given by
$D_\mu \Phi = (\partial_\mu -i g W^j_\mu \frac{\sigma^j}{2} - i
g^\prime B_\mu \frac{1}{2}) \Phi$. 

The lowest dimension operator that leads to quartic interactions but
does not exhibit two or three weak gauge boson vertices is dimension
8. The counting is straight foward: when can get a weak boson field
either from the covariant derivative of $\Phi$ or from the field
strength tensor. In either case the vector field is accompanied by a
VEV or a derivative. Therefore genuine quartic vertices are of
dimension 8 or higher.

There are three classes of such operators:

\subsubsection{ Operators containing just $D_\mu\Phi$}

The two independent operators in this class are
\begin{eqnarray}
  {\cal L}_{S,0} &=& \left [ \left ( D_\mu \Phi \right)^\dagger
 D_\nu \Phi \right ] \times
\left [ \left ( D^\mu \Phi \right)^\dagger
D^\nu \Phi \right ]
\\
  {\cal L}_{S,1} &=& \left [ \left ( D_\mu \Phi \right)^\dagger
 D^\mu \Phi  \right ] \times
\left [ \left ( D_\nu \Phi \right)^\dagger
D^\nu \Phi \right ]
\end{eqnarray}

\subsubsection{ Operators containing $D_\mu\Phi$ and field strength}

The operators in this class are:
\begin{eqnarray}
 {\cal L}_{M,0} &=&   \hbox{Tr}\left [ \hat{W}_{\mu\nu} \hat{W}^{\mu\nu} \right ]
\times  \left [ \left ( D_\beta \Phi \right)^\dagger
D^\beta \Phi \right ]
\\
 {\cal L}_{M,1} &=&   \hbox{Tr}\left [ \hat{W}_{\mu\nu} \hat{W}^{\nu\beta} \right ]
\times  \left [ \left ( D_\beta \Phi \right)^\dagger
D^\mu \Phi \right ]
\\
 {\cal L}_{M,2} &=&   \left [ B_{\mu\nu} B^{\mu\nu} \right ]
\times  \left [ \left ( D_\beta \Phi \right)^\dagger
D^\beta \Phi \right ]
\\
 {\cal L}_{M,3} &=&   \left [ B_{\mu\nu} B^{\nu\beta} \right ]
\times  \left [ \left ( D_\beta \Phi \right)^\dagger
D^\mu \Phi \right ]
\\
  {\cal L}_{M,4} &=& \left [ \left ( D_\mu \Phi \right)^\dagger \hat{W}_{\beta\nu}
 D^\mu \Phi  \right ] \times B^{\beta\nu}
\\
  {\cal L}_{M,5} &=& \left [ \left ( D_\mu \Phi \right)^\dagger \hat{W}_{\beta\nu}
 D^\nu \Phi  \right ] \times B^{\beta\mu}
\\
  {\cal L}_{M,6} &=& \left [ \left ( D_\mu \Phi \right)^\dagger \hat{W}_{\beta\nu}
\hat{W}^{\beta\nu} D^\mu \Phi  \right ] 
\\
  {\cal L}_{M,7} &=& \left [ \left ( D_\mu \Phi \right)^\dagger \hat{W}_{\beta\nu}
\hat{W}^{\beta\mu} D^\nu \Phi  \right ] 
\end{eqnarray}

\subsubsection{Operators containing just the field strength tensor}

The following operators containing just the field strength tensor 
also lead to quartic anomalous couplings:

\begin{eqnarray}
 {\cal L}_{T,0} &=&   \hbox{Tr}\left [ \hat{W}_{\mu\nu} \hat{W}^{\mu\nu} \right ]
\times   \hbox{Tr}\left [ \hat{W}_{\alpha\beta} \hat{W}^{\alpha\beta} \right ]
\\
 {\cal L}_{T,1} &=&   \hbox{Tr}\left [ \hat{W}_{\alpha\nu} \hat{W}^{\mu\beta} \right ]
\times   \hbox{Tr}\left [ \hat{W}_{\mu\beta} \hat{W}^{\alpha\nu} \right ]
\\
 {\cal L}_{T,2} &=&   \hbox{Tr}\left [ \hat{W}_{\alpha\mu} \hat{W}^{\mu\beta} \right ]
\times   \hbox{Tr}\left [ \hat{W}_{\beta\nu} \hat{W}^{\nu\alpha} \right ]
\\
 {\cal L}_{T,3} &=&   \hbox{Tr}\left [ \hat{W}_{\alpha\mu}
   \hat{W}^{\mu\beta}  \hat{W}^{\nu\alpha} \right ]
\times   B_{\beta\nu}
\\
 {\cal L}_{T,4} &=&   \hbox{Tr}\left [ \hat{W}_{\alpha\mu}
   \hat{W}^{\alpha\mu}  \hat{W}^{\beta\nu} \right ]
\times   B_{\beta\nu}
\\
 {\cal L}_{T,5} &=&   \hbox{Tr}\left [ \hat{W}_{\mu\nu} \hat{W}^{\mu\nu} \right ]
\times   B_{\alpha\beta} B^{\alpha\beta}
\\
 {\cal L}_{T,6} &=&   \hbox{Tr}\left [ \hat{W}_{\alpha\nu} \hat{W}^{\mu\beta} \right ]
\times   B_{\mu\beta} B^{\alpha\nu} 
\\
 {\cal L}_{T,7} &=&   \hbox{Tr}\left [ \hat{W}_{\alpha\mu} \hat{W}^{\mu\beta} \right ]
\times   B_{\beta\nu} B^{\nu\alpha} 
\\
 {\cal L}_{T,8} &=&   B_{\mu\nu} B^{\mu\nu}  B_{\alpha\beta} B^{\alpha\beta}
\\
 {\cal L}_{T,9} &=&  B_{\alpha\mu} B^{\mu\beta}   B_{\beta\nu} B^{\nu\alpha} 
\end{eqnarray}


\begin{thebibliography} {99}
%
\bibitem{anomalous} For a review see H.~Aihara {\it et al.}, 
in {\it Electroweak Symmetry
  Breaking and New Physics at the TeV Scale}, edited by T.~Barklow,
  S.~Dawson, H.~Haber and J.~Seigrist, (World Scientific,
  Singapore, 1996), p.\ 488 [arXiv:hep-ph/9503425].
%
\bibitem{lep} P.~Achard {\it et al.}  [L3 Collaboration],
  Phys.\ Lett.\ B {\bf 586}, 151 (2004);
P.~Abreu {\it et al.}  [DELPHI
  Collaboration], Phys.\ Lett.\ B {\bf 502}, 9 (2001);
  S.~Schael {\it et al.}  [ALEPH Collaboration],
  Phys.\ Lett.\ B {\bf 614} (2005) 7;
  G.~Abbiendi {\it et al.}  [OPAL Collaboration],
  Eur.\ Phys.\ J.\ C {\bf 33}, 463 (2004).
%
\bibitem{exp:LEP}
ALEPH, DELPHI, L3, OPAL and the LEP Electroweak Working Group,
arXiv:hep-ex/0511027.
%
\bibitem{teva} 
K.~Gounder [CDF Collaboration], arXiv:hep-ex/9903038;
B.~Abbott {\it et al.} [D\O\ Collaboration], Phys.\ Rev.\ D {\bf 62},
052005 (2000).
%
\bibitem{aew}
  C.~Arzt, M.~B.~Einhorn and J.~Wudka,
  Nucl.\ Phys.\ {\bf B433}, 41 (1995).
%
\bibitem{our:quartic} O.~J.~P.~\'Eboli, M.~C.~Gonzalez-Garcia, 
S.~M.~Lietti and S.~F.~Novaes, Phys.\ Rev.\ D {\bf 63}, 075008 (2001).
%
\bibitem{stir2} P.~J.~Dervan, A.~Signer, W.~J.~Stirling and
A.~Werthenbach , J.\ Phys.\ {\bf G26}, 607 (2000).
%
\bibitem{our:quartic2} O.~J.~P.~\'Eboli, M.~C.~Gonzalez-Garcia and 
S.~M.~Lietti,
  Phys.\ Rev.\ D {\bf 69}, 095005 (2004).
%
\bibitem{Belanger:1999aw}
G.~Belanger, F.~Boudjema, Y.~Kurihara, D.~Perret-Gallix and A.~Semenov,
Eur.\ Phys.\ J.\ C {\bf 13}, 283 (2000).
%
\bibitem{nlc}
G.~B\'elanger and F.~Boudjema, Phys.\ Lett.\ B {\bf 288}, 201 (1992); 
W.~J.~Stirling and A.~Werthenbach, Eur.\ Phys.\ J.\ C {\bf 14}, 103 (2000).
%
\bibitem{bel:bou}
G.~B\'elanger and F.~Boudjema, Phys.\ Lett.\ B {\bf 288}, 210 (1992).
%
\bibitem{ggnos}O.~J.~P.~\'Eboli, M.~B.~Magro, P.~G.~Mercadante
 and S.~F.~Novaes, Phys.\ Rev.\ D {\bf 52}, 15 (1995).
%
\bibitem{our:vvv} O.~J.~P.~\'Eboli, M.~C.~Gonzalez-Garcia, and
  S.~F.~Novaes, Nucl.\ Phys.\ {\bf B411}, 381 (1994).
%
\bibitem{bagger0} J.~Bagger, S.~Dawson and G.~Valencia,
Nucl.\ Phys.\ {\bf B399}, 364 (1993).
%
\bibitem{bagger} J.~Bagger  {\it et al.}, Phys.\ Rev.\ D {\bf
49}, 1246 (1994); Phys.\ Rev.\ D {\bf 52}, 3878
(1995).
%
\bibitem{dobado} A.~Dobado, D.~Espriu and M.~J.~Herrero, Z.\
Phys.\ C {\bf 50}, 205 (1991); A.~Dobado and M.~T.~Urdiales,
Z.\ Phys.\ C {\bf 17}, 965 (1996); A.~Dobado, M.~J.~Herrero,
E.~Ruiz, M.~T.~Urdiales and R.~Pelaez, Phys.\ Lett.\ B {\bf
352}, 400 (1995).
%
\bibitem{Belyaev:1998ih}
  A.~S.~Belyaev, O.~J.~P.~\'Eboli, M.~C.~Gonzalez-Garcia, J.~K.~Mizukoshi, 
S.~F.~Novaes and I.~Zacharov,
  Phys.\ Rev.\ D {\bf 59}, 015022 (1999).
%
\bibitem{equ} J.~M.~Cornwall, D.~N.~Levin and G.~Tiktopoulos, 
Phys.\ Rev.\ D {\bf 10}, 1145 (1974); C.~E.~Vayonakis, 
Lett.\ Nuovo Cim.\ {\bf 17}, 383 (1976); B.~W.~Lee, C.~Quigg and 
H.~B.~Thacker, Phys.\ Rev.\ D {\bf 16}, 1519
(1977); M.~S.~Chanowitz and M.~K.~Gaillard, Nucl.\ Phys.\
{\bf B261}, 379 (1985).
%
\bibitem{evb} G.~L.~Kane, W.~W.~Repko and W.~B.~Rolnick,
Phys.\ Lett.\ B {\bf 148}, 367 (1984); S.~Dawson, Nucl.\ Phys.\
{\bf B249}, 42 (1985).
%

\bibitem{unit} B.~Lee, C.~Quigg and H.~Thacker, Phys.\ Rev.\
Lett.\ {\bf 38}, 883 (1977); Phys.\ Rev.\ D {\bf 16}, 1519 (1977);
D.~Dicus and V.~Mathur, Phys.\ Rev.\ D {\bf 7}, 3111 (1973).
%
\bibitem{Barbieri:2004qk}
R.~Barbieri, A.~Pomarol, R.~Rattazzi and A.~Strumia,
Nucl.\ Phys.\ {\bf B703}, 127 (2004).
%
\bibitem{Brunstein:1996fz}
  A.~Brunstein, O.~J.~P.~\'Eboli and M.~C.~Gonzalez-Garcia,
  Phys.\ Lett.\ B {\bf 375}, 233 (1996).
%
\bibitem{linear} W.~Buchm\"uller and D.~Wyler, Nucl.\ Phys.\
{\bf B268}, 621 (1986); C.~J.~C.~Burges and H.~J.~Schnitzer, 
Nucl.\ Phys.\ {\bf B228}, 454 (1983); C.~N.~Leung,
S.~T.~Love and S.~Rao, Z.\ Phys.\ C {\bf 31}, 433 (1986); A.~De
R\'ujula, M.~B.~Gavela, P.~Hern\'andez and E.~Mass\'o, Nucl.\
Phys.\ {\bf B384}, 3 (1992); K.~Hagiwara, S.~Ishihara, R.~Szalapski and 
D.~Zeppenfeld, Phys.\ Lett.\ B {\bf 283}, 353 (1992); 
Phys.\ Rev.\ D {\bf 48}, 2182 (1993).
%
\bibitem{Appelquist}
  T.~Appelquist and C.~Bernard, Phys.\ Rev.\ D {\bf 22}, 200 (1980);
  A.~Longhitano, Phys.\ Rev.\ D {\bf 22}, 1166 (1980);  Nucl.\
  Phys.\ {\bf B188}, 118 (1981).
%
\bibitem{cgg} M.~S.~Chanowitz, M.~Golden and H.~Georgi, Phys.\ Rev.\
D {\bf 36}, 1490 (1987).
%
\bibitem{abc}
G.~Altarelli and R.~Barbieri, Phys.\ Lett.\  B {\bf 253}, 161 (1991);
G.~Altarelli, R.~Barbieri and S.~Jadach, Nucl.\ Phys.\  {\bf B369}, 3 (1992)
[Erratum-ibid.\ {\bf B376}, 444 (1992)].
%
\bibitem{pt}
M.~E.~Peskin and T.~Takeuchi, Phys.\ Rev.\ D {\bf 46}, 381 (1992).
%

\bibitem{mad} T.~Stelzer and W.~F.~Long, Comput.\ Phys.\ Commun.\ {\bf
    81}, 357 (1994).
%
\bibitem{helas} H.~Murayama, I.~Watanabe and K.~Hagiwara, KEK
report 91-11 (unpublished).
%
\bibitem{CMS} 
 A.~Denner, S.~Dittmaier, M.~Roth and D.~Wackeroth,
  Nucl.\ Phys.\ B {\bf 560}, 33 (1999).
%
\bibitem{CMS2}
  C.~Oleari and D.~Zeppenfeld,
  Phys.\ Rev.\ D {\bf 69}, 093004 (2004).
%
\bibitem{dieternew}
  B.~Jager, C.~Oleari and D.~Zeppenfeld,
  arXiv:hep-ph/0604200;
  B.~Jager, C.~Oleari and D.~Zeppenfeld,
  arXiv:hep-ph/0603177.
%
\bibitem{CTEQ5_pdf} H.~L.~Lai {\it et al.}  [CTEQ Collaboration],
  Eur.\ Phys.\ J.\ C {\bf 12}, 375 (2000).
%
\bibitem{boos} E.~E.~Boos, H.~J.~He , W.~Kilian, A.~Pukhov and
  P.~M.~Zerwas,  Phys.\ Rev.\ D {\bf 57}, 1553 (1997).

\bibitem{unit2} V.~Barger {\em et al.}, Phys.\ Rev.\ D {\bf 42},
3052 (1990).

\bibitem{veto}V.~Barger, R.~Phillips and D.~Zeppenfeld,
  Phys.\ Lett.\ B {\bf 346}, 106 (1995).
%
\bibitem{CZ} H.~Chehime and D.~Zeppenfeld, Phys.\ Rev.\ D {\bf 47},
  3898 (1993);
  D.~Rainwater, R.~Szalapski and D.~Zeppenfeld, Phys.\ Rev.\ D {\bf 54},
  6680 (1996).
%
\bibitem{rainth}
D.~Rainwater, Ph.D. thesis, report arXiv:hep-ph/9908378.
%
\bibitem{atlas}
V.~Cavasinni, D.~Costanzo and I.~Vivarelli, ATL-PHYS-2002-008;
S.~Asai {\it et al.},
Eur.\ Phys.\ J.\ C {\bf 32S2}, 19 (2004).
%
\bibitem{dieter}
 D.~L.~Rainwater and D.~Zeppenfeld,
  Phys.\ Rev.\ D {\bf 60}, 113004 (1999)
  [Erratum-ibid.\ D {\bf 61}, 099901 (2000)];
  N.~Kauer, T.~Plehn, D.~L.~Rainwater and D.~Zeppenfeld,
  Phys.\ Lett.\ B {\bf 503}, 113 (2001).
%
\bibitem{estback}
 See, for instance,  N.~Kauer,
  Phys.\ Rev.\ D {\bf 70}, 014020 (2004);
  C.~Buttar {\it et al.},
  arXiv:hep-ph/0604120;
  O.~J.~P.~\'Eboli and D.~Zeppenfeld,
  Phys.\ Lett.\ B {\bf 495}, 147 (2000).
%
\bibitem{nwafull}
  N.~Kauer,
  Phys.\ Rev.\ D {\bf 67}, 054013 (2003);
%
  N.~Kauer and D.~Zeppenfeld,
  Phys.\ Rev.\ D {\bf 65}, 014021 (2002).
%
\bibitem{grinstein}
  J.~Distler, B.~Grinstein and I.~Z.~Rothstein,
  arXiv:hep-ph/0604255.
\end{thebibliography}
\end{document}